\documentclass[%
 aip, jcp,
 amsmath,amssymb,
preprint,%
]{revtex4-1}

\usepackage{graphicx}
\usepackage{dcolumn}
\usepackage{bm}

\usepackage[utf8]{inputenc}
\usepackage[T1]{fontenc}
\usepackage{mathptmx}
\usepackage{etoolbox}
\usepackage{braket}
\usepackage{comment}

\makeatletter
\def\@email#1#2{%
 \endgroup
 \patchcmd{\titleblock@produce}
  {\frontmatter@RRAPformat}
  {\frontmatter@RRAPformat{\produce@RRAP{*#1\href{mailto:#2}{#2}}}\frontmatter@RRAPformat}
  {}{}
}%
\makeatother
\begin{document}


\title{Efficient Optimization of Low-Rank Antisymmetric Product of Geminals Wavefunction Using the Direct Givens Rotation Method}
\author{Airi Kawasaki}
\email{a\_kawasaki@gunma-u.ac.jp}
\thanks{These authors contributed equally to this work.}
\affiliation{ 
Division of Electronics and Mechanical Engineering, Graduate School of Science and Technology, Gunma University, 1-5-1 Tenjin-cho, Kiryu-shi, Gunma 376-8515, Japan
}%

\author{Rei Oshima}%
\thanks{These authors contributed equally to this work.}
\affiliation{ 
Department of Chemistry and Biochemistry, School of Advanced Science and Engineering, Waseda University, 3-4-1 Okubo, Shinjuku, Tokyo 169-8555, Japan
}%

\author{Naoki Nakatani}
\affiliation{%
Department of Chemistry, Graduate School of Science, Tokyo Metropolitan University, 1-1 Minami-Osawa, Hachioji-shi, Tokyo 192-0397, Japan
}%

\author{Hiromi Nakai}
\affiliation{ 
Department of Chemistry and Biochemistry, School of Advanced Science and Engineering, Waseda University, 3-4-1 Okubo, Shinjuku, Tokyo 169-8555, Japan
}%
\affiliation{%
Waseda Research Institute for Science and Engineering, Waseda University, 3-4-1 Okubo, Shinjuku, Tokyo 169-8555, Japan
}%

\date{\today}

\begin{abstract}
In our previous study, we proposed the low-rank antisymmetric product of geminals (APG) method, which reconstructs the wavefunction by extracting only the important eigenvalues from the APG wave function. However, its practical application was limited by the high computational cost from an orbital optimization process, making higher-rank calculations difficult. In this work, we reformulate the orbital part of the wavefunction using Givens rotation matrices, enabling an analytical treatment of the variational optimization. By combining the low-rank APG with the direct Givens rotation (DGR) method, we achieved a significant improvement in optimization efficiency. We applied the developed method to small molecular systems and confirmed that it provides high accuracy, while also significantly reducing the computational time compared to conventional methods.
\end{abstract}

\maketitle

\section{Introduction}
Since the development of density functional theory (DFT) \cite{1964PhRv..136..864H, kohn1965self}, a wide range of electronic structure calculations for atoms and molecules has been performed, leading to the rapid advancement of the field of quantum chemistry. However, because DFT constructs the reference system based on a single-particle approximation of the electron density, it faces limitations in accurately describing electron–electron interactions. Although DFT is widely used in most electronic structure calculations, there is still no reliable and efficient method for accurately treating strongly correlated electronic systems, in which the Coulomb repulsion between electrons plays a dominant role. Therefore, the development of accurate computational methods for strongly correlated electron systems remains a significant challenge.

To perform accurate calculations for strongly correlated electronic systems, geminal theories \cite{shull1959natural, 1965JMP.....6.1425C}, which describe electron pairs, have attracted increasing attention. Over the years, numerous geminal-based approaches have been introduced.

A geminal can be expressed using an antisymmetric matrix $F$ and creation operators, as follows:
\begin{eqnarray}
\hat{F}[k]\equiv \sum_{ab}^{2K} F[k]_{ab} \hat{c}^{\dag}_a\hat{c}^{\dag}_b
\end{eqnarray}
here, $2K$ represents the number of spin orbitals, and $k$ labels the type of geminal.
One of the representative geminal-based theories, the antisymmetric product of geminals (APG) wavefunction \cite{kutzelnigg1964w, mcweeny1963density}, can be written as a product of different geminals as follows:
\begin{eqnarray}
\ket{\Psi_{\mathrm{APG}}}= \hat{F}[1]\cdots\hat{F}[N/2]\ket{0}  \label{apg}
\end{eqnarray}
where $N$ is the number of electrons. 
Unlike the antisymmetrized geminal power (AGP) wavefunction \cite{1965JMP.....6.1425C}, which uses identical geminal products, the APG wavefunction employs different geminal products, allowing it to incorporate inter-geminal correlations. However, the use of different geminal products leads to difficulties in variational optimization and a high computational cost. As a result, many approximate wavefunctions based on APG have been developed \cite{arai1960theorem, hurley1953molecular, Tokmachev:2016:PPG, tarumi2013accelerating, jeszenszki2015local, mcweeny1980molecular, silver1969natural, johnson2017strategies, nicely1971geminal, johnson2013size, limacher2013new, johnson2017strategies}. Alternatively, wave functions that incorporate additional correlation into AGP have also been investigated \cite{henderson2019geminal, henderson2020correlating, dutta2020geminal, dutta2021construction, dutta2024correlated}.

In our previous work \cite{kawasaki2025low}, we overcame the variational difficulties of APG by transforming it into a linear combination of AGP wavefunctions using polynomial decomposition \cite{fischer1994sums}, which enabled applications to the Hubbard model and small molecular systems. Since the problem of high computational cost remains, we analyzed the optimized APG wavefunctions to explore approximate methods that capture the essential features of electron correlation, and found that each geminal coefficient matrix possesses approximately $N/2$ non-zero eigenvalues. Motivated by this finding, we developed a low-rank APG method that reconstructs the wavefunction using only the significant eigenvalues.

Low-rank APG is a flexible method that allows adjustment of the included electron correlation by changing the number of retained eigenvalues, that is, the rank. While this method has yielded favorable results in benchmark systems such as the Hubbard model, its applicability to higher ranks is severely limited by the intrinsic challenges of variational optimization, as will be discussed in the next section. Therefore, in this paper, we present a new formulation that employs the direct Givens rotation (DGR) method to enable calculations for larger systems and higher ranks.

This paper is organized as follows. Section \ref{s2} reviews the low-rank APG framework and the DGR method, and then outlines the combined formulation. Section \ref{s3} presents numerical results demonstrating the performance of the proposed approach. Conclusion is provided in Section \ref{s4}.

\section{Theory} \label{s2}
\subsection{Low-rank APG wavefunction}
In this section, we review the low-rank APG method.
The antisymmetric geminal coefficient matrix $F$ can be block-diagonalized into a banded real form using the Schur decomposition. 
\begin{eqnarray}
F=USU^* , \label{usu}
\end{eqnarray}
where $U$ is a $2K \times 2K$ unitary matrix and $S$ is
\begin{eqnarray}
S = \begin{pmatrix}
0 & \lambda^{(1)} &  &  & 0 \\
-\lambda^{(1)} & 0 & \ddots & &  \\
 & \ddots & \ddots & & \\
& & & 0 & \lambda^{(K)} \\
0 & & & -\lambda^{(K)} & 0
\end{pmatrix} \label{diag}.
\end{eqnarray}
Note that the eigenvalues of the antisymmetric matrix $F$ are purely imaginary, of the form $\lambda i$.

In our previous study \cite{kawasaki2025low}, we found that the number of non-zero eigenvalues of the antisymmetric geminal matrices used to construct the geminal wavefunctions is approximately $N/2$. Therefore, we developed the low-rank APG method, which reconstructs the AGP wavefunction using only the significant eigenvalues. The number of eigenvalues included corresponds to the rank of the matrix, and thus we refer to the wavefunctions as rank-1 APG, rank-2 APG, and so on, depending on the number of eigenvalues used.

In the rank-1 APG, for example, the $k$-th geminal operator can be approximated as
\begin{eqnarray}
\hat{F}[k] &=& \sum_{ab}^{2K}F[k]_{ab} \hat{c}^{\dag}_{a}\hat{c}^{\dag}_{b}\nonumber  \\
 &\rightarrow& \sum_{ab}^{2K}\frac{\lambda[k]^{(1)}}{2}\left(U[k]_{a1}U^{*}[k]_{\bar{1}b} - U[k]_{a\bar{1}}U^{*}[k]_{1b} \right) \hat{c}^{\dag}_{a}\hat{c}^{\dag}_{b} \nonumber \\
 &\equiv& \lambda[k]^{(1)} \hat{a}[k]^{\dag}_1 \hat{a}[k]^{\dag}_{\bar{1}} \nonumber \\
&\rightarrow& \lambda[k]^{(1)} \hat{a}^{\dag}_1 \hat{a}^{\dag}_{\bar{1}}  \label{rank-1x}
\end{eqnarray}
where we defined the $2K$ spin orbital index by $\{1,\bar{1},2,\bar{2},\ldots,K,\bar{K}\}$. In the third line of Eq.~(\ref{rank-1x}), a new operator is defined, and in the fourth line, an approximation is introduced in which the unitary transformations in all geminal coefficient matrices are assumed to be identical as follows.
\begin{eqnarray}
\sum_{a}^{2K} U[k]_{a1} \hat{c}^{\dag}_{a} \equiv \hat{a}[k]^{\dag}_{1} \rightarrow  \hat{a}^{\dag}_{1}
\end{eqnarray}
Using this geminal operator, the rank-1 APG wavefunction can be written as follows:
\begin{eqnarray}
\ket{\Psi_{\mathrm{r1APG}} } =\prod_{k}^{N/2} \lambda[k]^{(k)} \hat{a}^{\dag}_k \hat{a}^{\dag}_{\bar{k}} \ket{0} 
\end{eqnarray}
In a similar way, by increasing the number of included eigenvalues, rank-2 APG, rank-3 APG, and so on can be constructed.
The full-rank APG (i.e., the rank-$K$ APG) can be represented in the following form, which is equivalent to the antisymmetrized product of interacting geminals (APIG) wavefunction \cite{silver1969natural, johnson2017strategies, nicely1971geminal}.
\begin{eqnarray}
\ket{\Psi_{\mathrm{r}K\mathrm{APG}} } = \prod_{i}^{N/2}\left( \sum_{k}^{K}\lambda[i]^{(k)} \hat{a}^{\dag}_k \hat{a}^{\dag}_{\bar{k}}  \right)\ket{0} = \ket{\Psi_{\mathrm{APIG}} }
\end{eqnarray}
Similarly, the creation and annihilation operators in the Hamiltonian can be rewritten as follows.
\begin{eqnarray}
\hat{H}&=& \sum_{pq}t_{pq}\hat{c}^\dag_p\hat{c}_q + \sum_{pqrs}v_{pqrs}\hat{c}^\dag_p\hat{c}^\dag_q \hat{c}_s\hat{c}_r \nonumber  \\
&=& \sum_{PQ}\sum_{pq}t_{pq}U_{pP}U_{qQ}\hat{a}^\dag_P\hat{a}_Q + \sum_{PQRS}\sum_{pqrs}v_{pqrs}U_{pP}U_{qQ}U_{rR}U_{sS}\hat{a}^\dag_P\hat{a}^\dag_Q \hat{a}_S\hat{a}_R \nonumber  \\
&\equiv& \sum_{PQ} T_{PQ}\hat{a}^\dag_P\hat{a}_Q + \sum_{PQRS}V_{PQRS} \hat{a}^\dag_P\hat{a}^\dag_Q \hat{a}_S\hat{a}_R \label{hamiltonian}
\end{eqnarray}
Note that, in the Hamiltonian, only $T$ and $V$ in the last line of Eq.~(\ref{hamiltonian}) depend on the unitary matrix $U$.

The low-rank APG wavefunction is optimized by alternately optimizing the eigenvalue components and the unitary matrix components.
To maintain unitarity during optimization, our earlier work \cite{kawasaki2025low} expressed the unitary matrices in the form $U = \exp(X)$, where $X$ is an antisymmetric matrix, and treated $X$ as the variational parameter.
However, since the analytical derivative of $X$ is difficult to compute, the variational optimization was performed using numerical differentiation, which significantly increased the computational cost.
To overcome this difficulty, we introduce a formulation using Givens rotation matrices, which will be presented in the following section.

\subsection{Direct Givens rotation method} 

In this section, we provide an overview of the DGR method.  
Originally developed as an efficient approach to optimize unitary matrices within the Hartree--Fock (HF) method \cite{oshima2025direct}, the DGR method formulates the 
wavefunction optimization as the minimization of electronic energy with respect to unitary transformations of molecular orbitals (MOs).
\begin{eqnarray}
  C = C^{(0)} U, \label{c0u}
\end{eqnarray}
where $C^{(0)}$ is an initial molecular orbital coefficient matrix. The unitary transformation in Eq.~(\ref{c0u}) is expressed as a product of Givens rotation matrices $G(i,a)$.
\begin{eqnarray}
  U= \prod\limits_{i,a}^{} G\left( {i,a} \right) \label{Gia}
\end{eqnarray}
where indices $i$ and $a$ run occupied and virtual MOs, respectively. the Givens rotation matrix, $G(i,a)$, has 1 and 0 as 
the diagonal and off-diagonal elements, respectively, except for the $ii$, $ia$, $ai$, and $aa$ elements of $\cos\theta_{ia}$, 
$-\sin\theta_{ia}$, $\sin\theta_{ia}$, and $\cos\theta_{ia}$, respectively, as follows.
\begin{eqnarray}
G\left( {i,a} \right) = \left( {\begin{array}{*{20}{c}}
  1& \cdots &0& \cdots &0& \cdots &0\\
   \vdots & \ddots & \vdots &{}& \vdots &{}& \vdots \\
  0& \cdots &{\cos {\theta _{ia}}}& \cdots &{ - \sin {\theta _{ia}}}& \cdots &0\\
   \vdots &{}& \vdots & \ddots & \vdots &{}& \vdots \\
  0& \cdots &{\sin {\theta _{ia}}}& \cdots &{\cos {\theta _{ia}}}& \cdots &0\\
   \vdots &{}& \vdots &{}& \vdots & \ddots & \vdots \\
  0& \cdots &0& \cdots &0& \cdots &1
  \end{array}} \right) \label{Gdef}
\end{eqnarray}
The rotation angle matrix, $\mathnormal{\Theta}$, with the elements {$\theta_{ia}$}, is optimized through iterative calculations.
\begin{eqnarray}
\mathnormal{\Theta}^{\left( {n + 1} \right)} = \mathnormal{\Theta}^{\left( n \right)} + \Delta \mathnormal{\Theta}^{\left( n \right)} \label{theta_update}
\end{eqnarray}
where the superscript $n$ denotes the iteration step, and $\Delta \mathnormal{\Theta}^{(n)}$ represents the displacement. The displacement is 
determined using an approximate Hessian matrix $\tilde{H}$ constructed by the quasi-Newton method, as follows.
\begin{eqnarray}
\Delta \mathnormal{\Theta}^{\left( n \right)} =  - {\alpha\left( {{\tilde{H}^{\left( n \right)}}} \right)^{ - 1}}{g^{\left( n \right)}}
\end{eqnarray}
where $\alpha$ denotes the step size, and $g$ represents the gradient of the Lagrangian $L$ with respect to $\mathnormal{\Theta}$ 
evaluated at $\mathnormal{\Theta}$ = $\mathnormal{\Theta}^{(n)}$
\begin{eqnarray}
g_{ia}^{\left( n \right)} = {\left. {\frac{{\partial {L}}}{{\partial \theta _{ia}^{}}}} \right|_{{\mathnormal{\Theta}}
 = {{\mathnormal{\Theta}}^{\left( n \right)}}}} \label{gia}
\end{eqnarray}
In DGR method, the gradient $g$ is computed based on the error back-propagation (EBP) approach.
We introduce an adnoter index \( l \) to label the orbital pairs \((i,a)\), defining the intermediate quantity \( C^{(l,n)} \) as the molecular orbitals 
obtained by applying the \( l \)-th Givens rotation to the initial orbitals \( C^{(0)} \) at iteration \( n \), namely \( l = \{1, 2, \ldots, N_{\mathrm{pair}}\} \) 
corresponds to the pairs between occupied and virtual molecular orbitals \(\{(1, N_{\mathrm{occ}}+1), (1, N_{\mathrm{occ}}+2), \ldots, (N_{\mathrm{occ}}, N_{\mathrm{MO}})\}\), respectively.
\begin{eqnarray}
{C^{\left( {l,n} \right)}} = {C}^{\left( {l - 1,n} \right)}{G}^{\left( {l,n} \right)}
\end{eqnarray}
where $G^{(l,n)}$ denotes the $l$-th Givens rotation constructed from $\mathnormal{\Theta}^{(n)}$. 
By applying the chain rule using $C^{(l,n)}$, Eq. (\ref{gia}) is finally transformed as follows
\begin{eqnarray}
g_l^{\left( n \right)} = \mathop{\mathrm{tr}}\nolimits \left[ A^{\left( n \right)} G^{\left( N_{\mathrm{pair}}, n \right)\dag} 
\cdots G^{\left( l + 1, n \right)\dag} \left( C^{\left( l, n \right)} B^{\left( l, n \right)} \right)^\dag \right] \label{g_final}
\end{eqnarray}
Here, \( B \) is a matrix with elements \(-1\) and \(1\) at the \(ia\) and \(ai\) positions, respectively, and zeros elsewhere, and \( A \) is defined as follows.
\begin{eqnarray}
A_{\mu i}^{(n)} = \frac{{\partial L}}{{\partial C_{\mu i}^{({N_{{\rm{pair}}}},n)}}} \label{defA}
\end{eqnarray}
where $\mu$ denotes atomic orbitals. By regarding \( C^{(l,n)} \) as the \( l \)-th layer node in a neural network, Eq. (\ref{g_final})  calculates the gradient by sequentially applying \( G^{(l,n)} \) 
to \( A \) from the output layer toward the input layer, in a manner analogous to EBP.

\subsection{DGR method for Low-rank APG}
In the application of the DGR method to the low-rank APG method, the unitary matrix to be optimized is defined by Eq.~(\ref{usu}). 
In this formulation, the unitary matrix $U$ is of dimension $2K \times 2K$, and the initial coefficient matrix $C^{(0)}$ is set to the $2K \times 2K$ 
identity matrix. It should be noted that the number of rotation pairs, $N_{\mathrm{pair}}$, is given by $2K(2K - 1)/2$.
\begin{eqnarray}
  U= \prod\limits_{P,Q}^{} G\left( {P,Q} \right) \label{GPQ}
\end{eqnarray}
The optimization of $\theta$ is carried out in the same manner as in the HF method, following Eqs.~(\ref{theta_update})--(\ref{defA}).
The Lagrangian 
\(L\) in Eq.~(\ref{defA}) is given by the following expression for the rank-$x$ APG method.
\begin{eqnarray}
L = \bra{ \Psi_{\mathrm{r}x\mathrm{APG} } } \hat{H}  \ket{  \Psi_{\mathrm{r}x\mathrm{APG} } }
\end{eqnarray}
Therefore, $A$ is expressed as follows.
\begin{align}
A_{pP}^{(n)} 
&= 2\sum\limits_Q \left\langle  \Psi_{\mathrm{r}x\mathrm{APG} }  \middle| \hat{a}_P^\dagger \hat{a}_Q \middle|  \Psi_{\mathrm{r}x\mathrm{APG} } \right\rangle \widetilde{t}_{pQ} \notag \\
&\quad + 4\sum\limits_{QRS} \left\langle  \Psi_{\mathrm{r}x\mathrm{APG} }  \middle| \hat{a}_P^\dagger \hat{a}_Q^\dagger \hat{a}_S \hat{a}_R \middle|  \Psi_{\mathrm{r}x\mathrm{APG} } \right\rangle \widetilde{v}_{pQRS}
\end{align}
where $\widetilde{t}_{pQ}$ and $\widetilde{v}_{pQRS}$ are defined as follows, respectively.
\begin{eqnarray}
{\widetilde t_{pQ}} = \sum\limits_q {{t_{pq}}} {U_{qQ}}
\end{eqnarray}
\begin{eqnarray}
{\widetilde v_{pQRS}} = \sum\limits_{qrs} {{v_{pqrs}}} {U_{qQ}}{U_{rR}}{U_{sS}}
\end{eqnarray}

\section{Demonstrative applications} \label{s3}

In this section, we present demonstrative applications of the low-rank APG methods to small molecules. The calculations were performed using the STO-6G basis set.
Integrals of molecular Hamiltonian were calculated using PySCF. 
The results were compared with the exact diagonalization results using the H$\Phi$ package \cite{kawamura2017quantum} and unrestricted Hartree-Fock (UHF) using the mVMC package \cite{misawa2018mvmc}.

\subsection{H$_{2}$O}

First, the H$_{2}$O molecule was calculated at its equilibrium geometry using the following Cartesian coordinates, O(0, 0, 0), H(-1.809, 0, 0), H(0.453549, 1.751221, 0) in Bohr units. 

Although various wavefunctions can be constructed with the same rank in the low-rank APG approach, our previous study \cite{kawasaki2025low} showed that wavefunctions with smaller orbital overlaps yield better results. Therefore, we employed wavefunctions with minimal orbital overlap. The explicit forms are provided in the Appendix \ref{app}.

Table \ref{time} shows a comparison of the computation time between the previous \cite{kawasaki2025low} and present methods. In the rank-2 APG calculation of H$_{2}$O, the previous method required more than one week to achieve convergence, while the present method completed the calculation in less than 10 minutes. These results demonstrate that the present method significantly improves the computational efficiency.

Table \ref{low-rank_h2o} shows the total energies computed using the exact solution, APG, low-rank APG, AGP, and UHF. In our previous study \cite{kawasaki2025low}, due to optimization difficulties, we were only able to perform calculations up to rank-3 APG. In the present work, with improved optimization, we extended the calculations to rank-5 APG, corresponding to rank-$(N/2)$ APG for H$_{2}$O. 

The results for rank-4 and rank-5 APG are nearly identical, indicating that most of the essential eigenvalues are already included by rank-4 APG, which corresponds to rank-$(N/2 - 1)$ APG. Our previous study on the 6-electron Hubbard model showed that the accuracy plateaued beyond rank-2 APG, suggesting that, with a well-chosen wavefunction form, rank-$(N/2 - 1)$ APG may already incorporate the most important eigenvalues.

Even when the full rank is used in the low-rank APG method, its accuracy does not reach that of the original APG wavefunction. This discrepancy arises from an approximation introduced in the low-rank formulation, where all geminal matrices are assumed to share the same unitary transformation. The impact of this approximation depends on the system, particularly on how different the unitary matrices of the geminals are from each other. For the specific analysis, see Appendix \ref{app1}.

Although there is an accuracy limitation due to the approximation that assumes identical unitary matrices, it is found that even rank-2 APG achieves sufficiently good accuracy, closer to the exact solution than UHF or AGP.

\begin{table}
\caption{Comparison of computation time for H$_{2}$O (STO-6G) using the rank-2 APG between the present and previous methods.}
{\begin{tabular}{lcc} \hline
Rank-2 APG (H$_{2}$O) & Present method & Previous method \\ \hline
Time & $<$ 10 min & $ >$ 1 week \\ \hline
\end{tabular}}
\label{time}
\end{table}

\begin{table}
\caption{Total energy and the percentage of correlation energy ($E_{c}=E-E^{\mathrm{HF}}$) of exact diagonalization, APG, low-rank APG, AGP and UHF in H$_{2}$O (STO-6G).}
{\begin{tabular}{lcc} \hline
 H$_{2}$O   &Total energy (Hartree) & $(E_c/E_{c}^{\mathrm{APG}})*100$ ($\%$)  \\ \hline
 Exact & -75.728706 &  - \\
 APG   & -75.728417 &   100  \\
 Rank-5 APG&  -75.721926 & 86.9480   \\
 Rank-4 APG& -75.721926  & 86.9479  \\
 Rank-3 APG& -75.721869  & 86.8322  \\
 Rank-2 APG& -75.720217 &  83.5108  \\
 AGP  & -75.702368  & 47.6194 \\
 UHF     & -75.678686 & 0  \\ \hline
\end{tabular}}
\label{low-rank_h2o}
\end{table}

\begin{figure}
\centering
{\resizebox*{10cm}{!}{\includegraphics{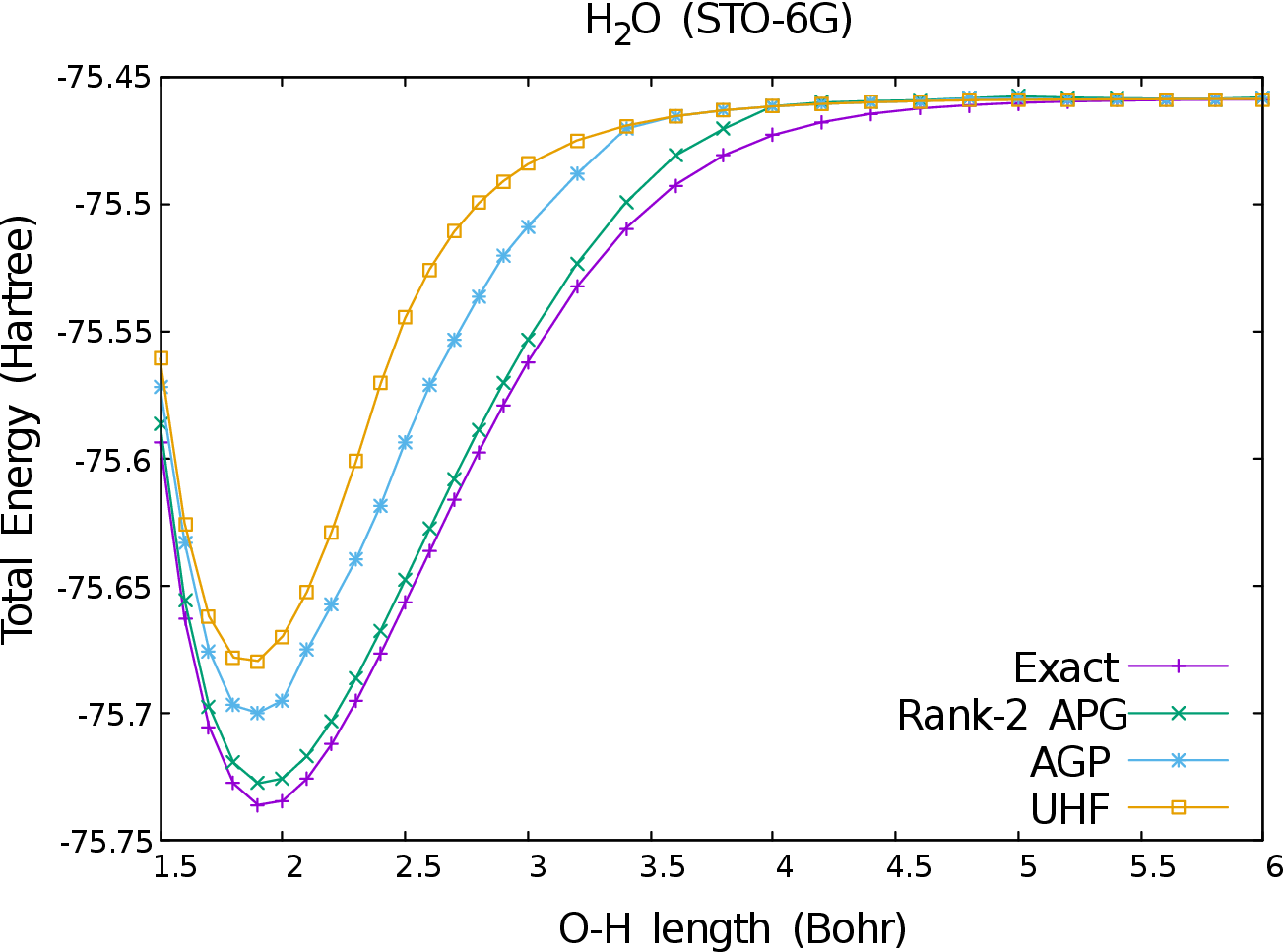}}}
\caption{Potential energy curves of H$_{2}$O (STO-6G).} 
\label{potential_h2o}
\end{figure}

Also, Figure \ref{potential_h2o} shows the potential energy curves calculated by the exact solution, rank-2 APG, AGP, and UHF. Here, the HOH bond angle was fixed at 104.5°, and the two O–H bond lengths were simultaneously elongated.

Across the entire range, rank-2 APG yields results close to the exact solution. While UHF and AGP perform well at long bond distances, their accuracy deteriorates at short bond distances compared to rank-2 APG. These results demonstrate that rank-2 APG is a highly accurate method capable of capturing both static and dynamic electron correlation.

\subsection{N$_{2}$}

Next, we increase the number of electrons and apply the rank-2 APG wavefunction to the N$_2$ molecule. See the Appendix \ref{app} for the form of the wavefunction.

\begin{figure}
\centering
{\resizebox*{10cm}{!}{\includegraphics{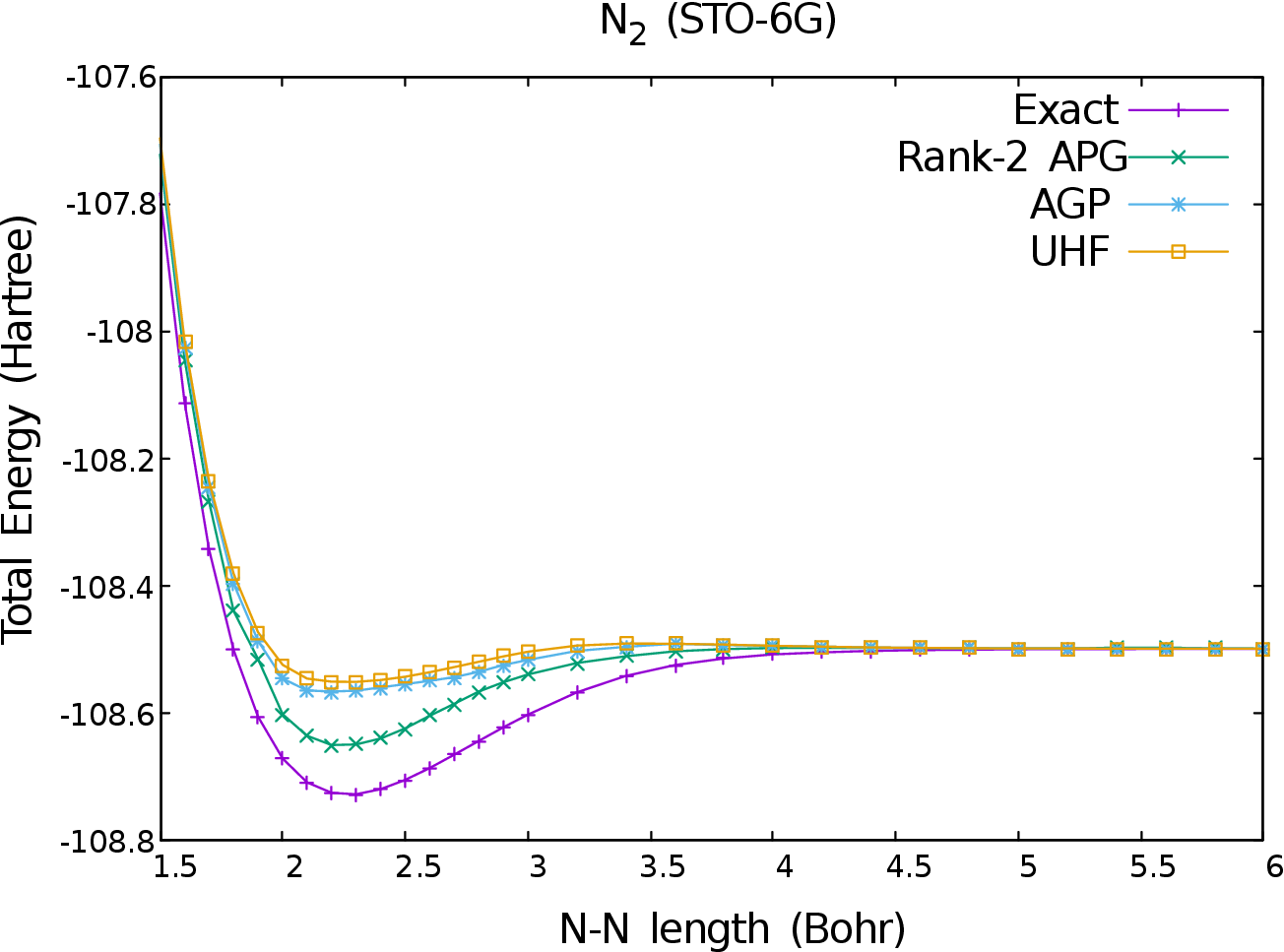}}}
\caption{Potential energy curves of N$_{2}$ (STO-6G).} 
\label{potential_n2}
\end{figure}

Figure \ref{potential_n2} presents the potential energy curves for the N$_2$.
As with H$_{2}$O, rank-2 APG also yields relatively accurate results for N$_{2}$ in the equilibrium bond distance region, where UHF and AGP struggle to describe the system properly.
Although it is well known that accurately describing the potential energy curve of N$_{2}$ is challenging due to the presence of strong static correlation, dynamic correlation effects, and the emergence of open-shell character in the dissociation limit, our results demonstrate that the rank-2 APG provides an accurate and reliable description across the entire dissociation process.

\section{Conclusion} \label{s4}

In this paper, we addressed the variational difficulties associated with the orbital optimization in the previous low-rank APG formulation, which relied on the parametrization $U = \exp(X)$, by employing Givens rotation matrices. Furthermore, we enhanced the efficiency of the optimization by combining our approach with the DGR method.

This approach significantly reduces the computational time, enabling us to handle higher ranks and systems with more electrons, and we applied the low-rank APG method to H$_2$O and N$_2$ to examine its accuracy. The low-rank APG method achieved higher accuracy than AGP and UHF even at rank-2, successfully capturing the potential energy curves of both H$_2$O and N$_2$. These results demonstrate that the low-rank APG method is capable of capturing both static and dynamic electron correlation.

Low-rank APG is a flexible method that allows tuning of the included electron correlation by varying the rank. Moreover, even at the same rank, it can represent a variety of wavefunction forms, providing additional flexibility. Therefore, it is expected that the low-rank APG framework can be tailored to construct wavefunctions adapted to specific systems, including, for example, those with localized electron correlation. We leave the exploration of such applications for future work.

\begin{acknowledgments}

Some calculations were performed at the Research Center for Computer Science (RCCS), Institute for Molecular Science (IMS), Okazaki, Japan (Project: 25-IMS-C124).
R.O. is grateful to JST SPRING, Grant Number JPMJSP2128.

\end{acknowledgments}

\section*{Conflict of Interest}
The authors have no conflicts to disclose.

\section*{Data Availability Statement}

The data that support the findings of this study are available within the article.

\appendix

\section{Explicit forms of the low-rank APG wavefunctions} \label{app}

The forms of the low-rank APG wavefunctions used in the H$_2$O calculations in this paper are as follows.
\begin{eqnarray}
\ket{\Psi_{\mathrm{r1APG}}} &=& \lambda[1]^{(1)} \hat{a}^{\dag}_{1} \hat{a}^{\dag}_{\bar{1}} \lambda[2]^{(2)} \hat{a}^{\dag}_{2} \hat{a}^{\dag}_{\bar{2}} \lambda[3]^{(3)} \hat{a}^{\dag}_{3} \hat{a}^{\dag}_{\bar{3}} \lambda[4]^{(4)} \hat{a}^{\dag}_{4} \hat{a}^{\dag}_{\bar{4}} \lambda[5]^{(5)} \hat{a}^{\dag}_{5} \hat{a}^{\dag}_{\bar{5}}  \ket{0} \\
\ket{\Psi_{\mathrm{r2APG}}} &=& \left(\lambda[1]^{(1)} \hat{a}^{\dag}_{1} \hat{a}^{\dag}_{\bar{1}} +\lambda[1]^{(6)} \hat{a}^{\dag}_{6} \hat{a}^{\dag}_{\bar{6}}  \right)\left(\lambda[2]^{(2)} \hat{a}^{\dag}_{2} \hat{a}^{\dag}_{\bar{2}} +\lambda[2]^{(7)} \hat{a}^{\dag}_{7} \hat{a}^{\dag}_{\bar{7}}  \right)\left(\lambda[3]^{(3)} \hat{a}^{\dag}_{3} \hat{a}^{\dag}_{\bar{3}} +\lambda[3]^{(1)} \hat{a}^{\dag}_{1} \hat{a}^{\dag}_{\bar{1}}  \right)  \nonumber \\
&&\left(\lambda[4]^{(4)} \hat{a}^{\dag}_{4} \hat{a}^{\dag}_{\bar{4}} +\lambda[4]^{(2)} \hat{a}^{\dag}_{2} \hat{a}^{\dag}_{\bar{2}}  \right)  
\left(\lambda[5]^{(5)} \hat{a}^{\dag}_{5} \hat{a}^{\dag}_{\bar{5}} +\lambda[5]^{(3)} \hat{a}^{\dag}_{3} \hat{a}^{\dag}_{\bar{3}}  \right) \ket{0} \\
\ket{\Psi_{\mathrm{r3APG}}} &=& \left(\lambda[1]^{(1)} \hat{a}^{\dag}_{1} \hat{a}^{\dag}_{\bar{1}} +\lambda[1]^{(6)} \hat{a}^{\dag}_{6} \hat{a}^{\dag}_{\bar{6}}+\lambda[1]^{(4)} \hat{a}^{\dag}_{4} \hat{a}^{\dag}_{\bar{4}}   \right)
\left(\lambda[2]^{(2)} \hat{a}^{\dag}_{2} \hat{a}^{\dag}_{\bar{2}} +\lambda[2]^{(7)} \hat{a}^{\dag}_{7} \hat{a}^{\dag}_{\bar{7}} +\lambda[2]^{(5)} \hat{a}^{\dag}_{5} \hat{a}^{\dag}_{\bar{5}}  \right) \nonumber  \\
&&\left(\lambda[3]^{(3)} \hat{a}^{\dag}_{3} \hat{a}^{\dag}_{\bar{3}} +\lambda[3]^{(1)} \hat{a}^{\dag}_{1} \hat{a}^{\dag}_{\bar{1}} +\lambda[3]^{(6)} \hat{a}^{\dag}_{6} \hat{a}^{\dag}_{\bar{6}}  \right)  
\left(\lambda[4]^{(4)} \hat{a}^{\dag}_{4} \hat{a}^{\dag}_{\bar{4}} +\lambda[4]^{(2)} \hat{a}^{\dag}_{2} \hat{a}^{\dag}_{\bar{2}} +\lambda[4]^{(7)} \hat{a}^{\dag}_{7} \hat{a}^{\dag}_{\bar{7}}   \right)  \nonumber \\
&&\left(\lambda[5]^{(5)} \hat{a}^{\dag}_{5} \hat{a}^{\dag}_{\bar{5}} +\lambda[5]^{(3)} \hat{a}^{\dag}_{3} \hat{a}^{\dag}_{\bar{3}} +\lambda[5]^{(1)} \hat{a}^{\dag}_{1} \hat{a}^{\dag}_{\bar{1}}  \right) \ket{0} \\
\ket{\Psi_{\mathrm{r4APG}}} &=& \left(\lambda[1]^{(1)} \hat{a}^{\dag}_{1} \hat{a}^{\dag}_{\bar{1}} +\lambda[1]^{(6)} \hat{a}^{\dag}_{6} \hat{a}^{\dag}_{\bar{6}}+\lambda[1]^{(4)} \hat{a}^{\dag}_{4} \hat{a}^{\dag}_{\bar{4}}  +\lambda[1]^{(2)} \hat{a}^{\dag}_{2} \hat{a}^{\dag}_{\bar{2}}  \right) \nonumber  \\
&&\left(\lambda[2]^{(2)} \hat{a}^{\dag}_{2} \hat{a}^{\dag}_{\bar{2}} +\lambda[2]^{(7)} \hat{a}^{\dag}_{7} \hat{a}^{\dag}_{\bar{7}} +\lambda[2]^{(5)} \hat{a}^{\dag}_{5} \hat{a}^{\dag}_{\bar{5}} +\lambda[2]^{(3)} \hat{a}^{\dag}_{3} \hat{a}^{\dag}_{\bar{3}}  \right) \nonumber  \\
&&\left(\lambda[3]^{(3)} \hat{a}^{\dag}_{3} \hat{a}^{\dag}_{\bar{3}} +\lambda[3]^{(1)} \hat{a}^{\dag}_{1} \hat{a}^{\dag}_{\bar{1}} +\lambda[3]^{(6)} \hat{a}^{\dag}_{6} \hat{a}^{\dag}_{\bar{6}} +\lambda[3]^{(4)} \hat{a}^{\dag}_{4} \hat{a}^{\dag}_{\bar{4}}  \right) \nonumber  \\
&&\left(\lambda[4]^{(4)} \hat{a}^{\dag}_{4} \hat{a}^{\dag}_{\bar{4}} +\lambda[4]^{(2)} \hat{a}^{\dag}_{2} \hat{a}^{\dag}_{\bar{2}} +\lambda[4]^{(7)} \hat{a}^{\dag}_{7} \hat{a}^{\dag}_{\bar{7}}  +\lambda[4]^{(5)} \hat{a}^{\dag}_{5} \hat{a}^{\dag}_{\bar{5}} \right)  \nonumber \\
&&\left(\lambda[5]^{(5)} \hat{a}^{\dag}_{5} \hat{a}^{\dag}_{\bar{5}} +\lambda[5]^{(3)} \hat{a}^{\dag}_{3} \hat{a}^{\dag}_{\bar{3}} +\lambda[5]^{(1)} \hat{a}^{\dag}_{1} \hat{a}^{\dag}_{\bar{1}} +\lambda[5]^{(6)} \hat{a}^{\dag}_{6} \hat{a}^{\dag}_{\bar{6}} \right) \ket{0} \\
\ket{\Psi_{\mathrm{r5APG}}} &=& \left(\lambda[1]^{(1)} \hat{a}^{\dag}_{1} \hat{a}^{\dag}_{\bar{1}} +\lambda[1]^{(6)} \hat{a}^{\dag}_{6} \hat{a}^{\dag}_{\bar{6}}+\lambda[1]^{(4)} \hat{a}^{\dag}_{4} \hat{a}^{\dag}_{\bar{4}}  +\lambda[1]^{(2)} \hat{a}^{\dag}_{2} \hat{a}^{\dag}_{\bar{2}} +\lambda[1]^{(7)} \hat{a}^{\dag}_{7} \hat{a}^{\dag}_{\bar{7}}  \right) \nonumber  \\
&&\left(\lambda[2]^{(2)} \hat{a}^{\dag}_{2} \hat{a}^{\dag}_{\bar{2}} +\lambda[2]^{(7)} \hat{a}^{\dag}_{7} \hat{a}^{\dag}_{\bar{7}} +\lambda[2]^{(5)} \hat{a}^{\dag}_{5} \hat{a}^{\dag}_{\bar{5}} +\lambda[2]^{(3)} \hat{a}^{\dag}_{3} \hat{a}^{\dag}_{\bar{3}} +\lambda[2]^{(1)} \hat{a}^{\dag}_{1} \hat{a}^{\dag}_{\bar{1}}  \right) \nonumber  \\
&&\left(\lambda[3]^{(3)} \hat{a}^{\dag}_{3} \hat{a}^{\dag}_{\bar{3}} +\lambda[3]^{(1)} \hat{a}^{\dag}_{1} \hat{a}^{\dag}_{\bar{1}} +\lambda[3]^{(6)} \hat{a}^{\dag}_{6} \hat{a}^{\dag}_{\bar{6}} +\lambda[3]^{(4)} \hat{a}^{\dag}_{4} \hat{a}^{\dag}_{\bar{4}} +\lambda[3]^{(2)} \hat{a}^{\dag}_{2} \hat{a}^{\dag}_{\bar{2}}  \right) \nonumber  \\
&&\left(\lambda[4]^{(4)} \hat{a}^{\dag}_{4} \hat{a}^{\dag}_{\bar{4}} +\lambda[4]^{(2)} \hat{a}^{\dag}_{2} \hat{a}^{\dag}_{\bar{2}} +\lambda[4]^{(7)} \hat{a}^{\dag}_{7} \hat{a}^{\dag}_{\bar{7}}  +\lambda[4]^{(5)} \hat{a}^{\dag}_{5} \hat{a}^{\dag}_{\bar{5}} +\lambda[4]^{(3)} \hat{a}^{\dag}_{3} \hat{a}^{\dag}_{\bar{3}} \right)  \nonumber \\
&&\left(\lambda[5]^{(5)} \hat{a}^{\dag}_{5} \hat{a}^{\dag}_{\bar{5}} +\lambda[5]^{(3)} \hat{a}^{\dag}_{3} \hat{a}^{\dag}_{\bar{3}} +\lambda[5]^{(1)} \hat{a}^{\dag}_{1} \hat{a}^{\dag}_{\bar{1}} +\lambda[5]^{(6)} \hat{a}^{\dag}_{6} \hat{a}^{\dag}_{\bar{6}} +\lambda[5]^{(4)} \hat{a}^{\dag}_{4} \hat{a}^{\dag}_{\bar{4}} \right) \ket{0}
\end{eqnarray}

The wavefunction used in the N$_2$ calculation is as follows.
\begin{eqnarray}
\ket{\Psi_{\mathrm{r2APG}}} &=& \left(\lambda[1]^{(1)} \hat{a}^{\dag}_{1} \hat{a}^{\dag}_{\bar{1}} +\lambda[1]^{(8)} \hat{a}^{\dag}_{8} \hat{a}^{\dag}_{\bar{8}}  \right)\left(\lambda[2]^{(2)} \hat{a}^{\dag}_{2} \hat{a}^{\dag}_{\bar{2}} +\lambda[2]^{(9)} \hat{a}^{\dag}_{9} \hat{a}^{\dag}_{\bar{9}}  \right)\left(\lambda[3]^{(3)} \hat{a}^{\dag}_{3} \hat{a}^{\dag}_{\bar{3}} +\lambda[3]^{(10)} \hat{a}^{\dag}_{10} \hat{a}^{\dag}_{\bar{10}}  \right)  \nonumber \\
&&\left(\lambda[4]^{(4)} \hat{a}^{\dag}_{4} \hat{a}^{\dag}_{\bar{4}} +\lambda[4]^{(1)} \hat{a}^{\dag}_{1} \hat{a}^{\dag}_{\bar{1}}  \right)  
\left(\lambda[5]^{(5)} \hat{a}^{\dag}_{5} \hat{a}^{\dag}_{\bar{5}} +\lambda[5]^{(2)} \hat{a}^{\dag}_{2} \hat{a}^{\dag}_{\bar{2}}  \right) \left(\lambda[6]^{(6)} \hat{a}^{\dag}_{6} \hat{a}^{\dag}_{\bar{6}} +\lambda[6]^{(3)} \hat{a}^{\dag}_{3} \hat{a}^{\dag}_{\bar{3}}  \right)  \nonumber \\
&& \left(\lambda[7]^{(7)} \hat{a}^{\dag}_{7} \hat{a}^{\dag}_{\bar{7}} +\lambda[7]^{(4)} \hat{a}^{\dag}_{4} \hat{a}^{\dag}_{\bar{4}}  \right)  \ket{0} 
\end{eqnarray}

\section{Effect of assuming identical unitary transformations} \label{app1}

In this appendix, we estimate the extent to which the approximation of using identical unitary transformations for all geminal coefficient matrices affects the results.

For example, in our previous study on the 6-site 6-electron Hubbard model \cite{kawasaki2025low}, the rank-$(N/2)$ APG recovered about $74\%$ of the APG correlation energy. In contrast, in the present H$_2$O calculations, it recovered about $87\%$. This shows that the accuracy of the low-rank APG varies depending on the system.

To evaluate how similar or different the unitary matrices are, we first optimized the APG wavefunction for each system. Then, we extracted the unitary matrices corresponding to the $N/2$ most important eigenvalues. For each pair of different geminals, we computed the matrix product $W = V{[k]} V^{\dag}{[l]} $ and calculated $\mathrm{tr}(W^{\dag} W)$.
The average values over all combinations of different geminal pairs are shown in Table \ref{apg_unitary}. Note that $\mathrm{tr}(W^{\dagger} W)$ equals $100\%$ when $k = l$, meaning the comparison is between the same geminal. These results indicate that the difference in overlap among the unitary matrices of different geminals plays a significant role in the accuracy of the low-rank APG method.

\begin{table}
\caption{Reproduction ratios of the APG correlation energy by rank-($N/2$) APG for the Hubbard model (6-site, 6-electron, $U/t = 10$) and H$_2$O (STO-6G), along with the average overlap of the unitary matrices in APG (normalized).}
{\begin{tabular}{lcc} \hline
  &Hubbard model &  H$_{2}$O  \\ \hline
$(E_{c}^{\mathrm{rank}-(N/2)\mathrm{APG}}/E_{c}^{\mathrm{APG}})*100$ ($\%$)  & 74.3571 &  86.9480\\
 $\mathrm{tr}(W^{\dag} W) *100$ ($\%$)   & 65.3584 &   81.5239  \\ \hline
\end{tabular}}
\label{apg_unitary}
\end{table}

\bibliography{cite}

\end{document}